\documentclass[aps,prl,twocolumn,showpacs,groupedaddress]{revtex4}
\usepackage[dvips]{graphicx}
\begin{document}
\title{Quantum partition noise of photo-created electron-hole pairs }
\author{L.-H. Reydellet}
\altaffiliation[Also at ]{Laboratoire de Physique de la Mati\`ere
Condens\'ee, Ecole Normale Sup\'erieure, Paris.}
\author{P. Roche}
\author{D. C. Glattli}
\altaffiliation[Also at ]{Laboratoire de Physique de la Mati\`ere
Condens\'ee, Ecole Normale Sup\'erieure, Paris.}
\affiliation{Service de Physique de l'Etat Condens\'e,\\
CEA Saclay, F-91191 Gif-Sur-Yvette, France}
\author{B. Etienne}
\author{Y. Jin}
\affiliation{Laboratoire de Photonique et de Nanostructures,\\
CNRS, Route de Nozay, F-91460 Marcoussis, France}
\date{\today}
\begin{abstract}
We show experimentally that even when no bias voltage is applied
to a quantum conductor, the electronic quantum partition noise can
be investigated using GHz radiofrequency irradiation of a
reservoir. Using a Quantum Point Contact configuration as the
ballistic conductor we are able to make an accurate determination
of the partition noise Fano factor resulting from the
photo-assisted shot noise. Applying both voltage bias and rf
irradiation we are able to make a definitive quantitative test of
the scattering theory of photo-assisted shot noise.
\end{abstract}

\maketitle

Can electron quantum partition noise be observed without net
electron transport? In this letter we investigate the shot noise
of a ballistic conductor under radiofrequency irradiation. We show
experimentally that photo-created electron hole pairs do generate
shot noise even when no net current flows through the conductor.

Current noise measurements associated with a d.c. current have
revealed many important phenomena in mesoscopic physics on the
last decade. Fundamental current fluctuations in a quantum
conductor out of equilibrium, called shot noise, are a sensitive
probe of both the charge and the statistics of the carriers
\cite{Dejong97,Blanter00PR336p1}. This approach has led to the
observation of the quantum suppression of shot noise in ballistic
conductors due to the Fermi statistics
\cite{Reznikov95PRL75p3340,Kumar96PRL76p2778}, the Laughlin
quasiparticle charge in the fractional quantum Hall regime
\cite{Saminadayar97PRL79p2526,Depicciotto97Nature389p162}, the
doubling of shot noise resulting from Andreev reflections at
Normal-Superconductor interfaces
\cite{Kozhevnikov00PRL84p3398,Jehl00Nature405p50} and the enhanced
noise due to multiple Andreev reflections
\cite{Dieleman97PRL79p3486,Cron01PRL86p4104}, the reduction factor
associated with the distribution of transmission probabilities in
diffusive or chaotic quantum conductors
\cite{Steinbach96PRL76p3806,Henny99PRB56p2871,Oberholzer02Naturep567,Schoelkopf98PRL80p2437}.

In general shot noise is produced when the quantum conductor is
driven out of equilibrium. In all the work quoted above, non
equilibrium was obtained by applying an electrochemical potential
difference between the electron reservoirs (or contacts) resulting
in a net current. Here, the origin is simple to understand.
Consider for simplicity a single mode quantum conductor at low
temperature with say the left reservoir biased by an
electrochemical potential difference $eV$ with respect to the
right. The left reservoir emits regularly electrons toward the
conductor at a frequency $eV/h$ as a result of the Fermi
statistics giving a incoming current $I_0=e(eV/h)$. If $D$ is the
electron transmission probability, the transmitted current
$I=DI_0$ gives the Landauer conductance $G=De^2/h$. As the regular
injection of electrons is noiseless, the only source of shot noise
corresponds to the quantum partition noise generated by electrons
either transmitted or reflected. The resulting current
fluctuations for a frequency bandwidth $\Delta f$ are $\overline{\Delta I^{2}%
}=2eI_0D(1-D)\Delta f$, where the term $D(1-D)$ is the variance of
the binomial statistics of the partitioning
\cite{Lesovik89JETPp592}.

Non equilibrium shot noise can however be produced when
\textit{no} voltage bias and hence \textit{no} mean current flows
through the conductor. For example, heating one reservoir is
expected to generate thermal noise but also shot noise with a
$D(1-D)$ dependence reflecting partitioning even when no
thermo-electric current is generated
\cite{Sukhorukov99PRB59p13054,Blanter00PR336p1}. Another non
equilibrium situation occurs when photons irradiate one side of
the quantum conductor (say the left one). This is the regime
addressed in this letter. To understand the mechanism, consider an
electron emitted from the left reservoir with an energy $\epsilon
\leq h\nu$ below the Fermi energy. The electron can be either
pumped to an energy $h\nu-\varepsilon$ above the Fermi energy with
probability $\mathcal{P}_1$ or unpumped with probability
$\mathcal{P}_0$. Unpumped electrons cannot generate current
fluctuations as the right reservoir also emits electrons at the
same energy. The Pauli principle imposes that both right and left
outgoing states be filled with one electron leading to no current
and hence no fluctuation. However the photo-pumped incoming
electrons and holes do generate noise: the right reservoir does
not emit electrons nor holes at energies $h\nu-\epsilon$ and
$-\epsilon$ respectively, so that partition noise is not
inhibited. The electron and hole incoming currents,
$I_0^{(e)}=\mathcal{P}_1 h\nu\,e/h$ and $I_0^{(h)}=-I_0^{(e)}$,
are sources of independent current fluctuations $\overline{\Delta
I^{(e)^2}}=2eI_0^{(e)}D(1-D) \Delta f$ and $\overline{\Delta
I^{(h)^2}}=\overline{\Delta I^{(e)^2}}$ respectively. They add
incoherently to give the total shot noise: $ \overline{\Delta
I^2}= 4 h\nu \frac{e^2}{h}D(1-D)\mathcal{P}_1\Delta f.$ In this
process the photon energy quantum $h\nu$ plays the role of the
bias voltage.

This is the basic shot noise mechanism generalized to multiple
modes and multiple photon absorption processes that we have
investigated experimentally. A complete formula for photo-assisted
shot noise, with and without voltage bias, has been derived in
\cite{Lesovik94PRL72p538,Pedersen98PRB58p12993}. For finite
voltage, the formula predicts a singularity in the shot noise
derivative at $eV=h\nu$ which was observed by the Yale group
\cite{Schoelkopf98PRL80p2437,Kozhevnikov00PRL84p3398} in a
diffusive sample, definitely showing the existence of
photo-assisted processes. The technique however measured the
derivative of the noise with bias and could not measure the full
shot noise in the zero current regime. Also it was not possible to
vary the transmission for an accurate test of the theory. Here, we
report \textit{total} noise measurements. We show that electrons
pumped to higher energy by photo-absorption do generate shot noise
when no bias voltage is applied between reservoirs (and hence no
current flows through the conductor). The sample, a Quantum Point
Contact (QPC), allows one to vary the transmission and fully
determine the Fano factor of the noise. In the doubly
non-equilibrium regime, where both rf and finite bias voltage are
applied, we recover the $eV=h\nu$ noise singularity providing
further evidence that photo-pumping is the basic underlying
mechanism.

The QPC is realized using a 2D electron gas in GaAs/AlGaAs with
$8\,10^{5}cm^{2}/Vs$ mobility and $4.8\,10^{11}cm^{-2}$ density.
Special etching of the mesa prior to evaporation of the QPC
metallic gates provides significant depletion at the QPC with zero
gate voltage. At low temperature, well defined conductance
plateaus for gate voltages ranging from $-55\,mV$ to $30\,mV$
allow accurately tuning of the transmission probability of the
first two modes. The noise measurements were performed using a
cross correlation technique \cite{Glattli97JAP81p7350} in the
$2.6$ to $4.2\,kHz$ range. The current noise power $S_{I}$ is
calculated from the voltage noise power $S_{V}$ measured across
the sample : $S_{I}=G^{2}S_{V}$ where $G$ is the differential
conductance recorded simultaneously. A $5.2\times10^{-28}
A^{2}/Hz$ background current noise results from the amplifier
current noise and the room temperature $100M\Omega$ current
source. The sensitivity of our experimental setup is checked
within $2\%$ both by measuring the quantum reduction of shot noise
\cite{Reznikov95PRL75p3340,Kumar96PRL76p2778} at transmission
$1/2$ and also by measuring the thermal noise for temperature
varying from $200\,mK$ to $600\,mK$. The base electronic
temperature is $94\pm5\,mK$ for a  $28\,mK$ refrigerator
temperature. The difference arises from the low loss coaxial cable
carrying the rf which brings a wide bandwidth high temperature
black body radiation to the sample \cite{Glattli97JAP81p7350}.

The first step of our experiment is to determine the frequencies
giving the highest coupling between the radiofrequency and the
sample. This is achieved by measuring a weak photocurrent which
never exceed $0.2\,nA$ in the explored rf power range (the
equivalent open circuit voltage is always lower than $k_{B}T$ such
that its effect on noise can be neglected in the experiments
described below). From this study we found a coupling sharply
peaked at two frequencies $17.32$ and $8.73\,GHz$ which therefore
will be used in the following.

It is interesting to compare the rf period with the transit time
of electrons between reservoirs. The distance between ohmic
contacts being $30\,\mu m$ and the elastic mean free path $9\,\mu
m$ we estimate the transit time to be $0.4\, ns$. This is shorter
than the coherence time and much longer than the rf period such
that application of photo-assisted model is legitimate.

\begin{figure}[h]
\includegraphics[width=8cm,keepaspectratio,clip]{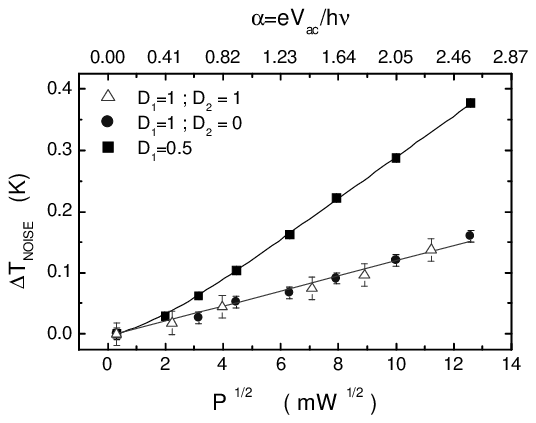}\caption{Excess
noise temperature as a function of the rf power $P$ on the top of
the fridge at $17.32\,GHz$. From the noise increase at
transmission 1, we deduced the electronic temperature increase due
to dissipation. The solid line for $D=0.5$ is a fit using
Eq.(\ref{Equi.eq}) when taking into account the temperature
increase. It gives the proportionality between $\alpha$ and
$P^{1/2}$}%
\label{Equi.fig}%
\end{figure}

We now present the results of our observation of photo-assisted
electron and hole partition noise with {\it no applied bias
voltage}. In the limit where $h\nu\gg k_{B}T$, the noise formula
\cite{Lesovik94PRL72p538,Pedersen98PRB58p12993,Schoelkopf98PRL80p2437}
is :
\begin{eqnarray}
T_{N} & = &T\,\left(
J_{0}^{2}(\alpha)+\frac{\sum_{n}D_{n}^{2}}{\sum_{n}D_{n}
}(1-J_{0}^{2}(\alpha))\right)\nonumber \\
& + & \sum_{l=1}^{+\infty}\frac{lh\nu}{k_B}
J_{l}^{2}(\alpha)\frac{\sum_{n}D_{n}(1-D_{n})}{\sum_{n}D_{n}}
\label{Equi.eq}
\end{eqnarray}
Here $D_{n}$ is the transmission probability of the $n^{th}$ mode,
$\alpha=eV_{ac}/h\nu$, $J_{l}$ the integer Bessel function of
order $l$ and $V_{ac}$ the rf voltage amplitude. The first term
represent the thermal noise of unpumped and pumped electrons. The
second term (which interests us here) is the partition noise of
photo-created electrons and holes scattered by the QPC as
discussed in the beginning ($\mathcal{P}_1=J_1^2$ here). When the
modes are either fully transmitted or reflected ( $D_{n}=1$ or $0$
), the noise is Johnson-Nyquist noise : $T_{N}=T$ and does not
depend on rf power. However, in a real experiments heating of the
reservoir by the rf power can not be excluded
\cite{Schoelkopf98PRL80p2437}. We thus first performed
measurements on the first conductance plateau ($D_{1}=G/G_{0}=1$)
and on the second plateau ($D_{1}=D_{2}=1$, $G/G_{0}=2$, where
$G_0=2e^2/h$). An increase of the noise is indeed observed when
increasing the rf power as shown in fig.(\ref{Equi.fig}). Here,
and in the following, the current noise power is expressed in
terms of noise temperature $T_{N}=S_{I}/4Gk_{B}$. Starting from a
base electron temperature of $94\,mK$, we observe a small noise
temperature increase which reaches $150\,mK$ for the highest power
used in the experiments. The increase is the same on the first and
second plateau. This indicates that heating occurs at the contact
and is not related to the physics of scattering at the QPC. It is
likely that heating of the contact results from rf absorption in
the lossy coaxial lines.
\begin{figure}[h]
\includegraphics[width=8cm,keepaspectratio,clip]{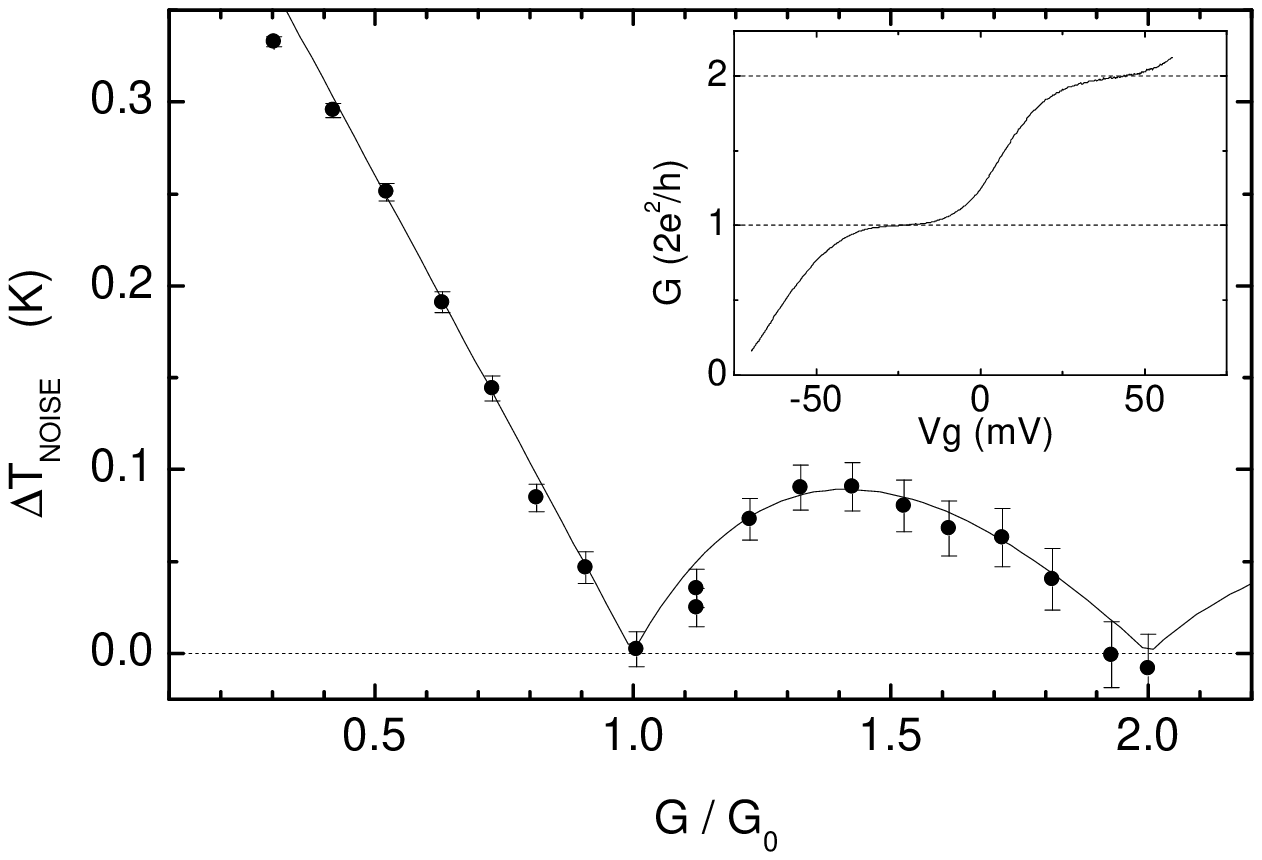}%
\caption{Noise temperature increase as a function of the
transmission $G/G_0$ when applying a $17.32\,GHz$ ac excitation
with $\alpha = 2.3$. The effect due to heating deduced from
Fig.(\ref{Equi.fig}) has been removed. The
solid line is the quantum suppression of the noise $\sum_{n} D_{n}%
(1-D_{n})/\sum_{n} D_{n}$. Inset: conductance versus gate voltage.}%
\label{Suppression.fig}%
\end{figure}

Having characterized the heating, we focus on the partition noise
regime expected for partial mode transmission.
Fig.(\ref{Equi.fig}) shows a much larger increase of the noise
temperature for $D_{1}=1/2$ than for $D_{1}=1$ and
$D_{1}=D_{2}=1$. Can this difference to be attributed to partition
noise of photo-pumped electrons or to thermally assisted shot
noise? It is straightforward to give a quantitative estimation of
the latter process. For an average temperature increase of the
left and right reservoirs $\Delta T=(\Delta T_{left}+\Delta
T_{right})/2$, one finds that the noise temperature increase never
exceeds $\Delta T_{N}^{th}\leq\Delta T+(1-D_{1})[2\ln2-1]\Delta
T$. According to the study at transmission 1 and 2, $\Delta T$ is
at most $150\,mK$, which gives $\Delta T_{N}^{th}\leq179\,mK$ at
half transmission, i.e. only $29\,mK$ above the noise temperature
increase observed on the plateaus. The much larger noise observed
at $D_{1}=1/2$ strongly suggests the presence of photo-assisted
process. Taking the heating into account Eq.(\ref{Equi.eq}) fits
the experimental results extremely well. From this we deduce
$\alpha=eV_{ac}/h\nu$ to be a function of the square root of the
applied rf power. This provides a calibration of the rf coupling
which will be used below \footnote{Referred to the power $P$
delivered by the rf source before attenuation, we find a
proportionality between $\alpha$ and $10^{P(dBm)/20}$ equal to
$0.204\pm0.004$ at $17.32\,GHz$ and $1.5\pm0.1$ at $8.73\,GHz$.}.

To fully characterize the partition noise of photo-pumped
electrons and holes a systematic study as a function of
transmission has been performed. The transmissions are measured
using simultaneous measurement of the conductance $G = G_0\sum_n
D_n$. Fig.(\ref{Suppression.fig}) shows the noise temperature
variation versus transmission for $\alpha=2.3$. The first term of
Eq.(\ref{Equi.eq}) has been subtracted from the data as the
transmissions $D_n$, $\alpha$ and the dependence of the electronic
temperature with rf power are known. This allows better comparison
with the second term of Eq.(\ref{Equi.eq}) which represents the
electron hole partition noise. The variation of the noise is
clearly proportional to the Fano factor
\cite{Reznikov95PRL75p3340,Kumar96PRL76p2778}
$\sum_{n}D_{n}(1-D_{n})/\sum_{n}D_{n}$ and unambiguously
demonstrates the photo-assisted partition noise. Indeed the solid
curve is the theoretical comparison with no adjustable parameter.
Fig.(\ref{Suppression.fig}) is the central result of this present
work. This is the first observation of electron hole partition
noise without net electron transport.
\begin{figure}[h]
\includegraphics[width=8cm,keepaspectratio,clip]{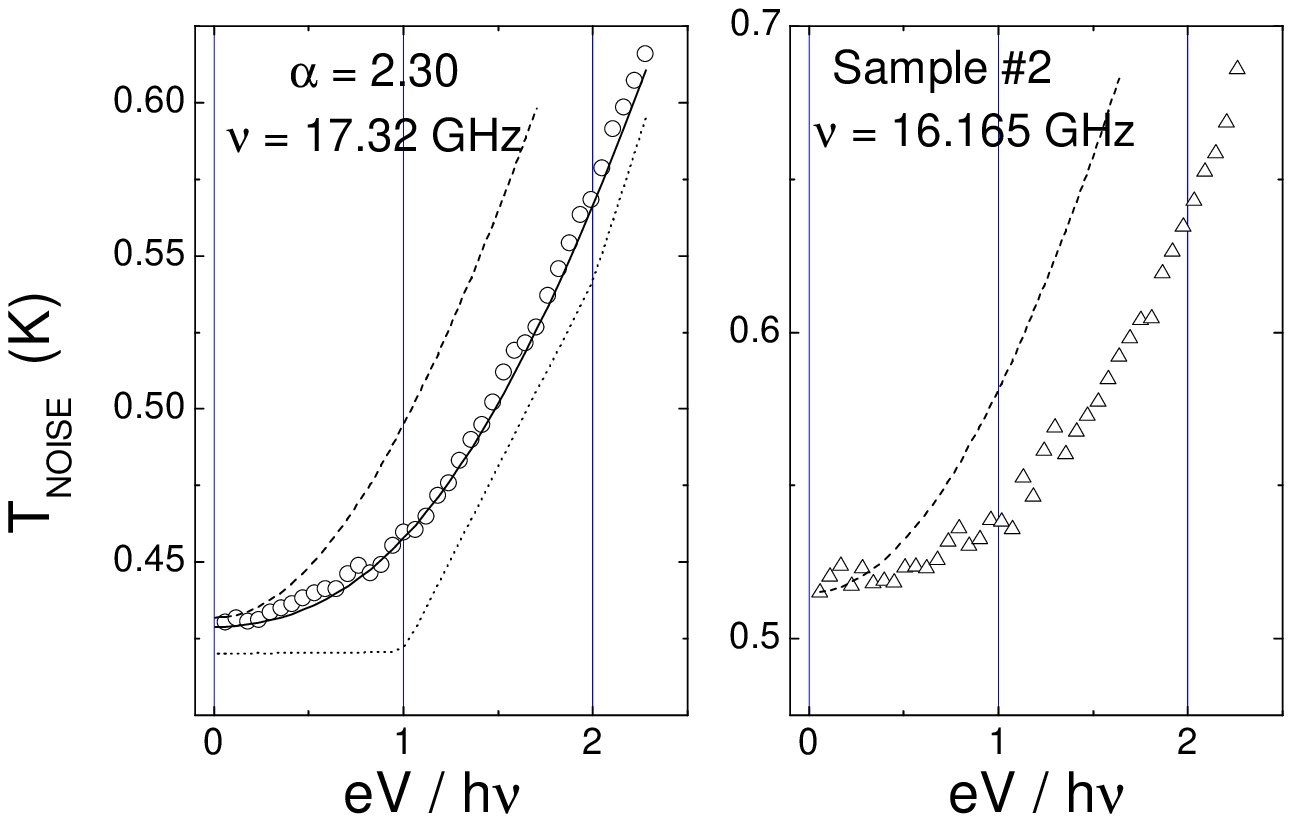}\caption{Left graph:
Noise temperature as a function of $eV/h\nu$ with 17.32 GHz ac
excitation. The measured electron temperature was $229\,mK$ and
$\alpha=2.3$. The dotted line is the expected photo-assisted noise
at zero temperature with $\alpha=2.3$ shifted for comparison. The
dashed line is the non photo-assisted shot noise with $T=430\,mK$.
Continuous line is the photo-assisted noise calculated using
Eq.(\ref{PAN}) without adjustable parameter. Right graph: Noise
temperature as a function of $eV/h\nu$ with 16.165 GHz ac
excitation for a different sample. The dashed line is the expected
non photo-assisted shot
noise for $T=520\,mK$.}%
\label{Cusp.fig}%
\end{figure}

A further check that photo-assisted noise is the basic underlying
mechanism is to apply both rf and finite bias voltage
simultaneously on the QPC.  It is known that a singularity occurs
in the shot noise variation at $eV=h\nu$
\cite{Lesovik94PRL72p538,Schoelkopf98PRL80p2437,Kozhevnikov00PRL84p3398,Pedersen98PRB58p12993}.
This is simple to understand. Lets assume that the left reservoir
has a chemical potential rise $eV$ above the right. The noise
arises from transport noise of pumped and unpumped electrons in
the energy window $eV$ and from photo-pumped electron hole pairs
in the energy window $h\nu-eV$. The latter process observed
previously at zero bias, disappears for $eV\ge h\nu$ leading to
the singularity. The complete formula including multiple photon
processes is \cite{Lesovik94PRL72p538}:
\begin{widetext}
\begin{equation}
T_N = T \frac{\sum_{n}D_{n}^{2}}{\sum_{n}D_{n}}+\frac{\sum_{n}
D_{n}(1-D_{n})}{\sum_{n}D_{n}}\sum_{\pm}\sum_{l=0}^{+\infty}J_{l}^{2}(\alpha)
\frac{eV\pm lh\nu}{2k_{B}}\coth\left(\frac{eV\pm
lh\nu}{2k_{B}T}\right) \label{PAN}
\end{equation}
\end{widetext}

\begin{figure}[ptb]
\includegraphics[width=8cm,keepaspectratio,clip]{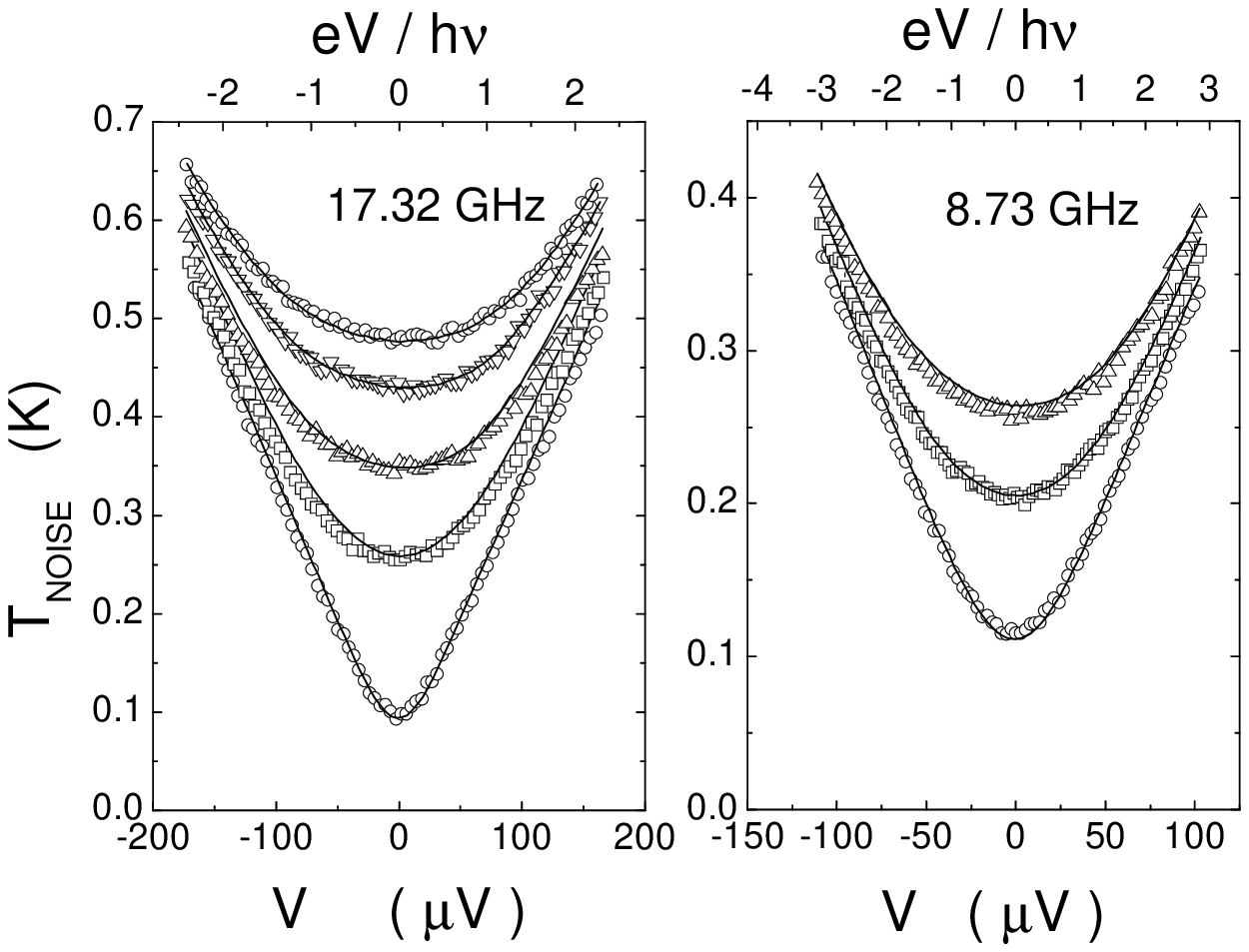}\caption{Noise
versus bias voltage. The solid lines are the theoretical curves
with $\alpha$ and $T$ deduced from the equilibrium noise under rf
illumination (see Fig.\ref{Equi.fig}). For the $17.32\,GHz$
curves, $\alpha =0.065;\,1.29;\,1.83;\,2.30;\,2.58$ and
$T=94;\,168;\,200;\,229;\,246mK$ and for the $8.73\,GHz$ curves,
$\alpha=0.51;\,1.81;\,2.56$ and
$T=105.5;\,145;\,167.7mK$.}%
\label{Total.fig}%
\end{figure}
Fig.(\ref{Cusp.fig}), left graph, shows total noise measurements
versus bias voltage at transmission $1/2$ for the same conditions
as Fig.(\ref{Suppression.fig})($\alpha=2.3$, $T=229\,mK$). As we
can see on Fig.(\ref{Cusp.fig}) for bias higher than $h\nu/e$ the
noise starts to increase more rapidly, a hallmark of
photo-assisted processes. This behavior cannot be attributed to
simple thermal rounding, even if we assume that the $V=0$ noise
temperature was corresponding to $430\,mK$ electronic temperature
(dashed line). In order to better reveal the expected
singularities in the voltage, we have plotted the noise at
$T\approx0$ (dotted curve). As all parameters $\alpha$, $T$, and
$D$ are known we can use Eq.(\ref{PAN}) to make a comparison with
our data. The agreement is excellent. Even the thermal rounding of
the singularity at $eV=h\nu$ is well reproduced. The singularity
of noise has been also observed using a different sample with a
slightly different coupling and pumping frequency and is displayed
on the right graph of Fig.(\ref{Cusp.fig}).

Finally Fig.(\ref{Total.fig}) shows a set of curves for various rf
power values at $17.32$ and $8.73\,GHz$ \footnote{The weak
photocurrent $I_{ph}$ gives an offset in $V$ which is less than
$6\mu V$ for $D=0.5$. This offset is removed.}. The curves show
that for the two different frequencies, the voltage scale of noise
variation is determined by the photon energy quantum $h\nu$ and
not by the thermal energy scale. From the regime of nearly pure
shot noise to the regime of strongly photo-assisted shot noise all
curves compares accurately with theory without any adjustable
parameter.

To summarize, absolute noise measurements on a quantum point
contact under rf irradiation have provided the first demonstration
that the quantum partition noise of electrons can be observed when
no current flows through the sample. This is possible because
photo created electron hole pairs scatters at the point contact
generating current fluctuations. The photo-assisted process has
been further brought into evidence when applying finite voltage
leading to singularities for $eV=h\nu$. All data show perfect
agreement with the quantum scattering theory of photo-assisted
shot noise.

\end{document}